\documentstyle[jkas,rotating]{article}
\beginpage{1}
\endpage{12}
\year{2009} \volume{41}

\runningauthor {Lim et al.} \year{2009} \volume{41}
\beginpage{1}\endpage{12}
\runningtitle{Standard Stars}
\begin{document}
\title{CCD Photometry of Standard Stars at Maidanak Astronomical Observatory in Uzbekstan: Transformations and Comparisons}
\author{Beomdu Lim$^{1}$, Hwankyung Sung$^1$, M. S. Bessell$^2$, R. Karimov$^3$, M. Ibrahimov$^3$}
\address{$^1$ Department of Astronomy and Space Science,
Sejong University,Seoul,Korea\\
 {\it e-mail}: bdlim1210@sju.ac.kr, sungh@sejong.ac.kr}
\address{$^2$ Research School of Astronomy \& Astrophysics, Australian National University,
Private Bag, Weston Creek PO, ACT 2611, Australia\\ {\it e-mail}: bessell@mso.anu.edu.au}
\address{$^3$ Ulugh Beg Astronomical Institute, 33 Astronomical Street, Tashkent 700052, Uzbekistan\\
{\it e-mail}: rivkat@astrin.uzsci.net, mansur@astrin.uzsci.net }

\address{\normalsize{\it ()}}
\offprints{H. Sung}
\abstract{Observation of standard stars is of crucial importance in stellar photometry. We have studied the
standard transformation relations of the $UBVRI$ CCD photometric system at the Maidanak Astronomical Observatory
in Uzbekistan. All observations were made with the AZT-22 1.5m telescope, SITe 2k CCD or Fairchild 486 CCD, and
standard Bessell $UBVRI$ filters from 2003 August to 2007 September. We observed many standard stars around the
celestial equator observed by SAAO astronomers.
The atmospheric extinction coefficients, photometric zero points, and time variation of photometric zero points
of each night were determined. Secondary extinction coefficients and photometric zero points were very stable,
while primary extinction coefficients showed a distinct seasonal variation. We also determined the transformation
coefficients for each filter. For $B$, $V$, $R$, and $I$ filters, the transformation to the SAAO standard system
could be achieved with a straight line or a combination of two straight lines. However, in the case of the $U$ filter and
Fairchild 486 CCD combination, a significant non-linear correction term - related to the size of Balmer jump or the
strength of the Balmer lines - of up to $0.08 \ mags$  was required.
We found that our data matched well the SAAO photometry in $V$,
$B-V$, $V-I$, and $R-I$. But in $U-B$, the difference in zero point was about $3.6 \ mmag$ and the
scatter was about $0.02 \ mag$. We attribute the relatively large scatter in $U-B$ to the larger error in $U$
of the SAAO photometry. We confirm the mostly small differences between the SAAO standard $UBVRI$ system and the Landolt
standard system. We also
attempted to interpret the seasonal variation of the atmospheric extinction coefficients in the context of scattering
 sources in the earth's atmosphere.}

  \maketitle

\section{INTRODUCTION}
Many photometric systems have been developed for various
 astronomical purposes since photoelectric photometry was introduced. A photometric system is defined by the
central wavelengths and width of its passbands and its set of standard stars. Photometric systems can be
divided into three categories according to the width of their passbands - broad, intermediate, and narrow band
system (Bessell 2006). The `Johnson-Cousins' $UBVR_CI_C$ system is the most widely used broad-band
photometric system.

The $UBV$ photometric system defined by Johnson and Morgan in 1953 was designed around the
 established MK spectral classification (Johnson \& Morgan 1953). They defined the zero point of
 all color indices to be those of the bright northern main-sequence star Vega with spectral type A0.
Subsequently they published several papers of photoelectric photometric data that comprises
the original $UBV$ photometric standard star lists.
 Most of these standard stars are very bright and in addition many variable stars were also included.
A few years later, Johnson (1966) extended his photometric system up to mid-IR.   The advent of newer and more
sensitive and reliable photoelectric detectors in the 1970s and CCDs in the 1990s opened accurate photometry to
fainter stars. The original  Johnson standards were inappropriate for modern photometry and nowadays
the Kron-Cousins $R_CI_C$ system is widely used instead of Johnson's $R_JI_J$ because of the more precise standards.

 Currently two sets of standard star regions are mainly used in $UBV(RI)_C$ photometry. The first is the
E-region standard stars centered at declination $-45\,^{\circ}$ in the southern hemisphere. The
other is the Landolt equatorial standard stars. The photoelectric $UBV$ photometry of E-region
standard stars were established by A. W. J. Cousins (1978). It is very closely tied to the Johnson UBV system.
The astronomers at the South Africa Astronomical Observatory (SAAO)
are continuously improving the accuracy and precision of the E-region stars.
The SAAO standard stars are known as the most precise system available currently.
However, there are several limitations in the E-region standard stars
- the majority are too bright, they lack
extremely blue or red stars, and are sparsely distributed.
SAAO observers have tried to overcome these limitations. Menzies et al. (1991) (hereafter M91) performed extensive
photometry of Landolt's equatorial standard stars as
program stars using E-region standard stars. Kilkenny et al. (1998) published additional photometric
data for very blue and red stars.

Landolt (1973, 1983, 1992) also published extensive lists of standard stars in the celestial equator.
Landolt (1973) was confined to $UBV$ only, but based on Cousins E-region stars. Later Landolt (1983) (hereafter L83)
extended his standard system to $(RI)_C$. Owing to the emergence of sensitive two-dimensional detectors, such as CCDs,
the necessity of
many faint standard stars covering a wide color range in a small field of view was obvious. Landolt
(1992) (hereafter L92) is the only available standard star list to meet these requirements. He also observed many
blue and red standard stars selected from various catalogues. Recently Landolt (2007)
published new standard stars at declination $\approx-50\,^{\circ}$.

 There are several reports on the systematic differences between SAAO and
Landolt systems, although Landolt's photometry were based on Cousins standards. Those systematic differences in $V-I, B-V$,
and $ U-B$ are discussed in
M91, Bessell (1995), and Sung \& Bessell (2000) (hereafter SB00). The differences in $(U-B)$ are sufficiently large that
they make it
difficult to get reliable physical quantities for early type stars. These systematic differences have arisen through
differences in instrumental system passbands and standardization problems associated with a lack of red and blue
E-region standards.

Precise standardized photometry cannot be obtained without a detailed knowledge of the photometric system, its
standard stars, and the characteristics of the observing site including atmospheric extinction corrections. To investigate
 the characteristics of the observing site as well as the photometric system of Maidanak Astronomical Observatory (MAO),
we observed many standard stars in the celestial
equator over the last five years with a Fairchild 486 CCD (hereafter 4k CCD) or a SITe 2000 $\times$ 800
 CCD (2k CCD), and Bessell $UBVRI$ filters (See Lim et al. 2008 for the
characteristics of 4k CCD system). Atmospheric extinction and standard transformation are dealt with in \S II.
Comparisons between our results and Landolt photometry are made in \S III. The characteristics of the atmospheric
extinction coefficients are discussed in \S IV. We also present variable star candidates from our
observations in that section. The summary is given in \S V.

\section{ATMOSPHERIC EXTINCTION AND STANDARD TRANSFORMATION}
\subsection{MAO CCD system}
All observations were made with the AZT-22 1.5m telescope (f/7.74), 4k CCD or 2k CCD, and Bessell $UBVRI$
filters at MAO. The characteristics of the 4k CCD system can be found in the previous paper (Lim et al. 2008).

\begin{figure}[!t]
\epsfxsize=9cm \epsfbox{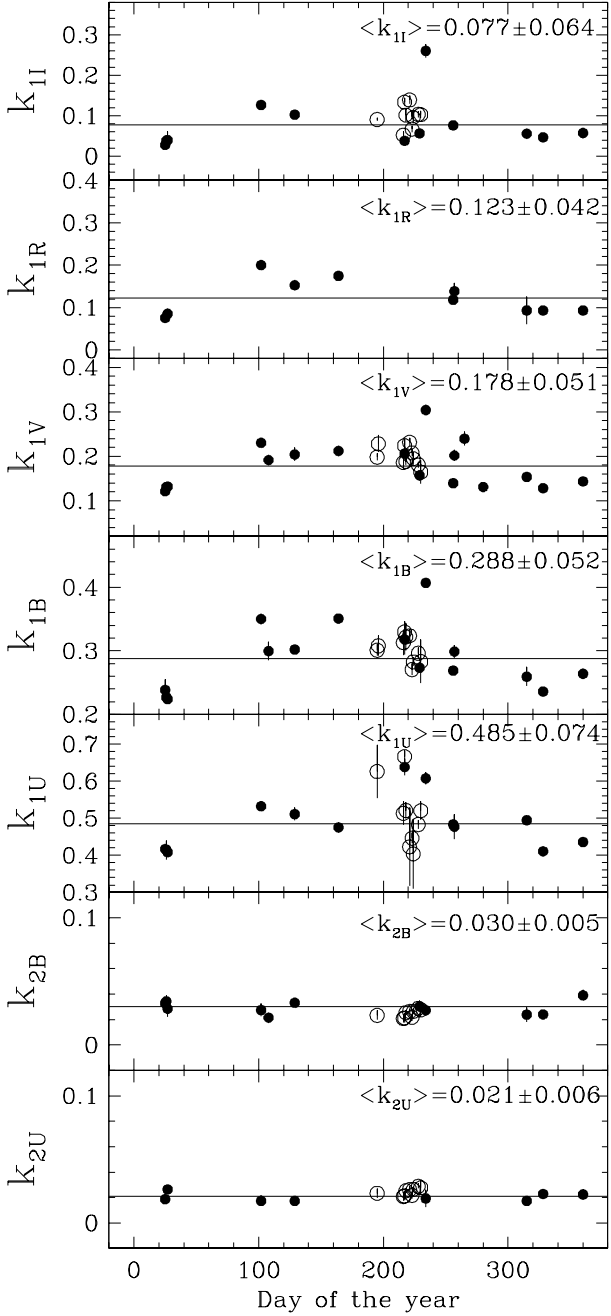} \caption{Extinction coefficients
for each filter. The line and numbers given at the upper right of each panel
indicates the mean value of atmospheric extinction coefficients.
The filled and open circles represent
the extinction coefficients obtained with the 4k CCD and 2k CCD,
respectively. All primary extinction coefficients show an obvious
seasonal variation, whereas secondary extinction coefficients are stable.}
\end{figure}

\begin{figure}[!]
\epsfxsize=9cm \epsfbox{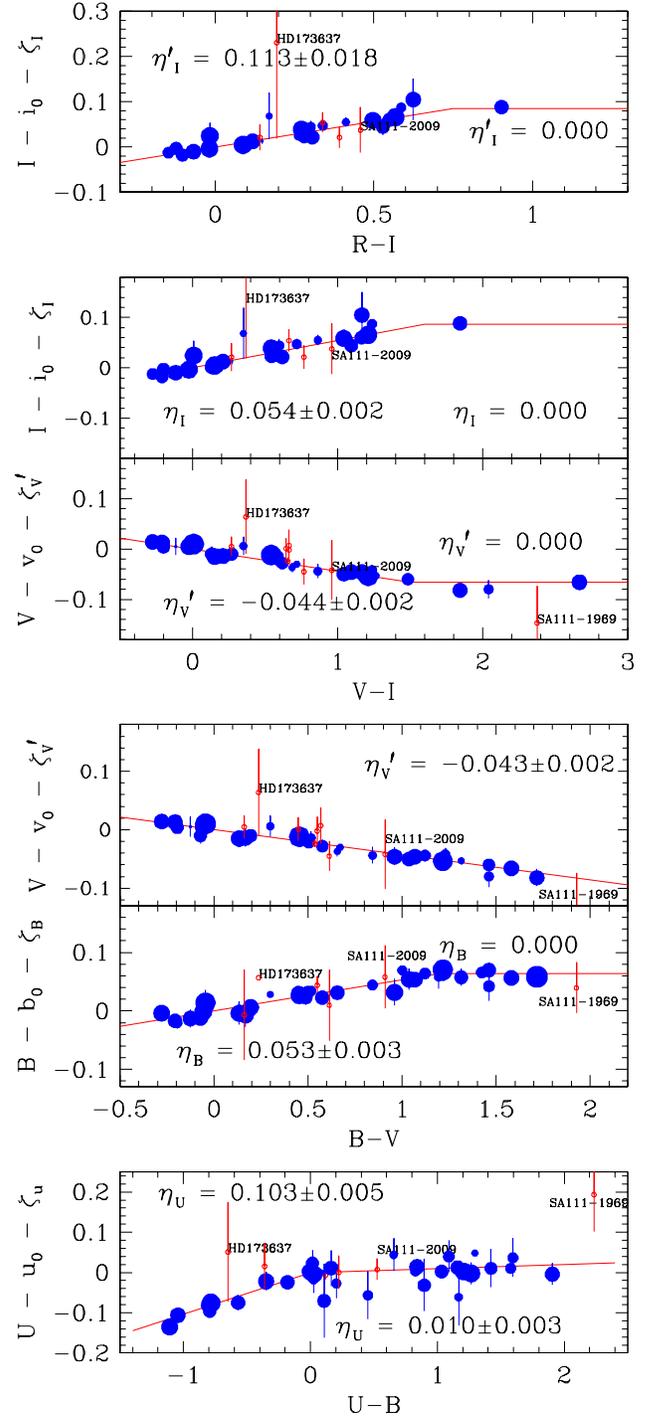} \caption{The standard
transformation relation for the SITe 2k CCD. The size of the circles is
proportional to the numbers of observations. The solid line represents
the transformation relation. Open circles denote suspected variables. The error bar is
the quadratic sum of the standard deviation in M91 or Kilkenny et al. (1998) and
our data.}
\end{figure}

\subsection{Observation and Data Reduction}
Our observations were mainly made from 2003 August to 2007 September.
Twenty five nights were photometric. A total of $2513$ measurements ($436$ in $I$, $225$ in $R$,
$663$ in $V$, $585$ in $B$, $604$ in $U$) of SAAO standard stars in M91 and
Kilkenny et al. (1998) were used to derive atmospheric extinction and
transformation coefficients. We also measured the magnitude of Landolt standard
stars in the observed images. To get quality data we used an appropriate exposure time
for a given standard star. Nevertheless some bright stars were
saturated even in short exposure images and some faint stars, especially in $U$, were too
faint to get reliable data.

Pre-processing was performed using the IRAF/ CCDRED package. All processes for the
4k CCD are described by Lim et al. (2008). After removing instrumental signatures,
we performed simple aperture photometry for standard stars. An aperture of $10{\arcsec}$ for the 2k CCD and
$14{\arcsec}$ for the 4k CCD was used for standard star photometry.
Standard stars with good signal-to-noise ratios ($\epsilon \leq 0.01$) were then used
to determine the extinction and transformation coefficients for each filter.

\subsection{Extinction Coefficient}
Atmospheric extinction is caused by absorption and scattering from gas molecules, dust particles and aerosols in the earth's
atmosphere. Its value depends primarily on airmass - the line-of-sight length passing through the earth's
atmosphere. In addition, since a component of the extinction varies with wavelength, the
value of the extinction measured across a wide filter passband will differ depending on the spectral energy
distribution of stars. We correct for this effect by looking for a primary or first extinction coefficient
that is dependent on airmass but is independent of color, and a secondary extinction coefficient that depends
also on color. The secondary extinction coefficient is negligible in $V$ or redder passbands. The magnitude
corrected for the atmospheric extinction is given by

\begin{equation}
m_{\lambda,0} = m_{\lambda} - (k_{1\lambda} - k_{2\lambda}C_0)X,
\end{equation}

\noindent
where $m_{\lambda0}$, $m_{\lambda}$, $k_{1\lambda}$, $k_{2\lambda}$, $C_0$, and $X$ are the extinction-
corrected magnitude, observed instrumental magnitude, primary extinction coefficient, secondary extinction coefficient,
relavant color index, and airmass, respectively. In general, we observed many
standard regions several times at various airmasses. To ensure a long baseline in airmass
we often observed standard stars near the meridian and again at zenith distances of $\approx60$.

Although we used four on-chip amplifiers for the 4k CCD to save CCD read-out time, we could find no
noticeable difference in photometric zero points associated with them. The atmospheric
extinction coefficients presented in Figure 1 and Table 1 for the 4k CCD are therefore the weighted mean values
of those determined from the standard stars in each quadrant. The secondary extinction coefficients are only
given for $U$ and $B$,  those in $V$, $R$ and $I$ were too small to be considered meaningful.
We plot the extinction coefficients against day of the year in Fig. 1. Those extinction
coefficients obtained with the 2k CCD agree well with those determined with the 4k CCD data.

\begin{table*}
\begin{center}
\scriptsize
\bf{\sc  Table 1.}\\
\sc{Extinction coefficients} \\
\begin{tabular}{cccccccc}
\\ \hline \hline
 Date of Obs   &   $k_{1I}$  &  $k_{1R}$  & $k_{1V}$  &  $k_{1B}$ &  $k_{1U}$ &  CCD &  Time variation\\
 \hline\\
2003.08.18.  & 0.1027 &   -    &  0.1646 & 0.2831 & 0.5204 & SITe 2000 $\times$ 800       & $\circ$ \\
2004.08.10.  & 0.0669 &   -    &  0.2078 & 0.2706 & 0.4463 & SITe 2000 $\times$ 800       & $\circ$ \\
2004.08.11.  & 0.0957 &   -    &  0.1947 & 0.2827 & 0.4038 & SITe 2000 $\times$ 800       & $\circ$ \\
2004.08.15.  & 0.1033 &   -    &  0.1810 & 0.2963 & 0.4830 & SITe 2000 $\times$ 800       & $\circ$ \\
2005.08.04.  & 0.0521 &   -    &  0.1859 & 0.3129 & 0.5134 & SITe 2000 $\times$ 800       & $\circ$ \\
2005.08.05.  & 0.1337 &   -    &  0.2241 & 0.3297 & 0.6663 & SITe 2000 $\times$ 800       & $\circ$ \\
2005.08.06.  & 0.1014 &   -    &  0.1885 & 0.3219 & 0.5210 & SITe 2000 $\times$ 800       &  \\
2005.08.09.  & 0.1391 &   -    &  0.2314 & 0.3238 & 0.4230 & SITe 2000 $\times$ 800       & $\circ$ \\
2006.07.14.  & 0.0905 &   -    &  0.1982 & 0.3010 & 0.6261 & SITe 2000 $\times$ 800       & $\circ$ \\
2006.07.15.  &   -    &   -    &  0.2282 & 0.3080 &   -    & SITe 2000 $\times$ 800       & $\circ$ \\
2006.08.17.  & 0.0563 &   -    &  0.1572 & 0.2734 &   -    & Fairchild 4096 $\times$ 4096 & $\circ$ \\
2006.08.22.  & 0.2601 &   -    &  0.3040 & 0.4069 & 0.6074 & Fairchild 4096 $\times$ 4096 &  \\
2006.11.11.  & 0.0556 & 0.0932 &  0.1536 & 0.2594 & 0.4948 & Fairchild 4096 $\times$ 4096 &  \\
2006.11.24.  & 0.0465 & 0.0931 &  0.1285 & 0.2360 & 0.4107 & Fairchild 4096 $\times$ 4096 &  \\
2006.12.26.  & 0.0571 & 0.0932 &  0.1436 & 0.2640 & 0.4358 & Fairchild 4096 $\times$ 4096 &  \\
2007.01.25.  & 0.0277 & 0.0757 &  0.1214 & 0.2384 & 0.4169 & Fairchild 4096 $\times$ 4096 &  \\
2007.01.26.  & 0.0386 &   -    &  0.1308 & 0.2269 & 0.4141 & Fairchild 4096 $\times$ 4096 &  \\
2007.01.27.  & 0.0395 & 0.0855 &  0.1320 & 0.2241 & 0.4085 & Fairchild 4096 $\times$ 4096 &  \\
2007.04.12.  & 0.1265 & 0.2000 &  0.2302 & 0.3501 & 0.5322 & Fairchild 4096 $\times$ 4096 &  \\
2007.04.18.  &   -    &   -    &  0.1916 & 0.2996 &   -    & Fairchild 4096 $\times$ 4096 &  \\
2007.05.09.  & 0.1024 & 0.1527 &  0.2042 & 0.3020 & 0.5110 & Fairchild 4096 $\times$ 4096 &  \\
2007.06.13.  &   -    & 0.1747 &  0.2120 & 0.3509 & 0.4753 & Fairchild 4096 $\times$ 4096 &  \\
2007.08.05.  & 0.0380 &   -    &  0.2060 & 0.3180 & 0.6380 & Fairchild 4096 $\times$ 4096 & $\circ$ \\
2007.09.13.  & 0.0762 & 0.1183 &  0.1394 & 0.2688 & 0.4832 & Fairchild 4096 $\times$ 4096 & $\circ$ \\
2007.09.14.  &   -    & 0.1386 &  0.2018 & 0.2989 & 0.4767 & Fairchild 4096 $\times$ 4096 &  \\
 mean        & 0.0862 & 0.1225 &  0.1864 & 0.2939 & 0.4958 &                              &          \\
\hline
\end{tabular}
\end{center}
\end{table*}

There are obvious seasonal variations in the primary extinction coefficients
in all filters. The coefficients are larger in summer, but smaller in winter. In addition,
the extinction coefficients in summer show a large scatter (see \S IV(a) for the discussion). The mean
extinction coefficients are slightly larger in $V$ and $R$ than those at Siding Spring Observatory (SSO), but
in $U$, $B$, and $I$ are slightly smaller (SB00). On the other hand, we
could not find any seasonal variations in the secondary extinction coefficients,
and therefore, observers at MAO can use mean values for the secondary extinction coefficient in $U$ and $B$.

\begin{figure*}[!t]
\epsfxsize=18cm \epsfbox{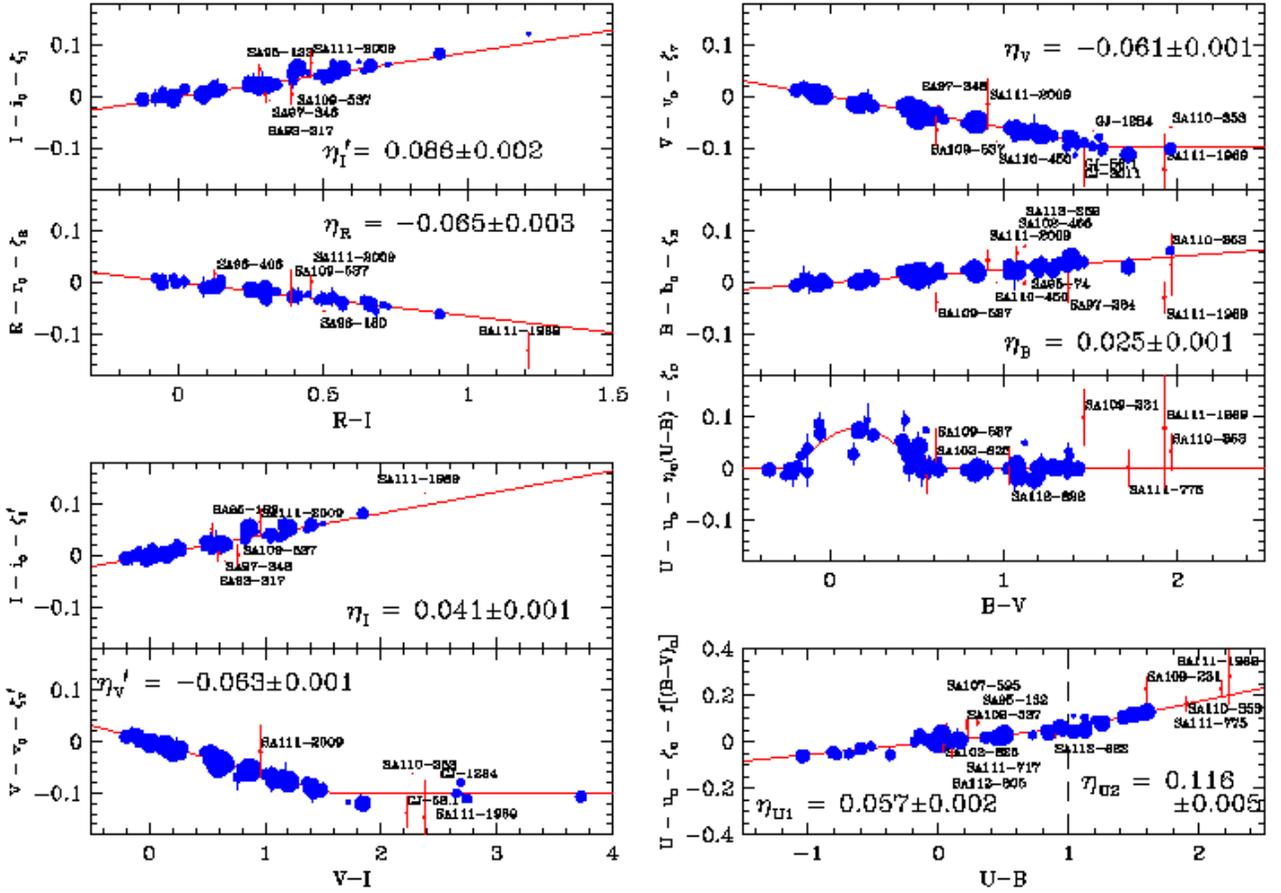} \caption{The standard
transformation relation for the Fairchild 486 CCD. The vertical dashline denotes the color
where the slope of the transformation changes. The other symboles are the same as Figure 2. }
\end{figure*}

\subsection{Transformation Coefficients}

Any photometric system is defined by the filters and detectors used in the observations.
A slight deviation between the standard magnitudes and atmospheric-extinction-corrected natural instrumental
magnitudes is expected, and needs to be corrected for to achieve the highest accuracy. Such differences are tracked
and corrected for through observation of many standard stars with as full a range of color as possible.
The correction term
between two systems is called the transformation coefficients, and is related as follow.

\begin{equation}
M_{\lambda} = m_{\lambda,0}+\eta_{\lambda}\cdot C_0+ \alpha_\lambda \cdot \hat{UT}+\zeta_{\lambda}
\end{equation}

\noindent
where $M_{\lambda}$, $m_{\lambda,0}$, $\eta_{\lambda}$, $C_0$, $\alpha$, $\hat{UT}$, $\zeta_{\lambda}$
represent the standard magnitude, atmospheric extinction-corrected instrumental magnitude as defined in equation (1),
transformation coefficient, relevant color index, time-variation coefficient, time difference relative
to midnight, and photometric zero point, respectively. All parameters are determined by using the weighted least square
method (weight $= weight \ factor/\epsilon^2$). The $weight \ factor$
is proportional to the reliability level of standard stars and the relative quality of each night.
$\epsilon$ is the photometric error.

The final transformation relations for the 2k CCD are shown in Figure 2, and these coefficients are listed in Table 2. All
transformations to the SAAO system are a straight line or combination of two straight lines. But the transformation
relation for very red stars ($V-I \geq 2.0$ or $R-I > 1.0$) is uncertain due to a lack of very red standard stars.
We also showed the transformation relation for the 4k CCD in Figure 3. For $B$, $V$, $R$, and $I$ filters,
the transformation to the SAAO standard system can be achieved with a straight line
or a combination of two straight lines. However, in $U$, there is a conspicuous non-linear
correction term related to the size of the Balmer jump or the strength of the Balmer lines. Sung et al. (1998) and SB00 also
introduced a similar non-linear correction term in the transformation of the SSO $U$ filter.
The maximum size of the non-linear correction term is up to $0.08 \ mag$.

The transmission function of the Bessell $U$ filter matches well with the
total response function of the original Johnson's $U$ filter and the quantum efficiency of
the photomultiplier tube. The non-linear correction term in the $U$ transformation must result from the effective wavelength
of the $U$ passband being pushed to far to the red. With most CCD photometry this results from a steep change in the
quantum efficiency of the CCD across the $U$ passband, but the average quantum efficiency of a Fairchild 486 CCD
chip (Fairchild homepage) is high and nearly flat in $U$  so the passband shift must result from some other optical
component cutting off the short wavelength side of the $U4$ response (e.g. dewar window).

The necessity of very red standard stars has increased recently as the finding of very red, low-mass objects
has become an important topic in astronomy. We tried to determine the transformation relation of $I$ and $R$
for very red colors, but failed because all the red standard stars were saturated even in the shortest exposure time.

\subsection{Time Variation of Photometric Zero Points}

The photometric zero points depend primarily on the light gathering power of the photometric system,
i.e. the size and state of the primary mirror and the quantum efficiency of the detectors. In addition,
changes in the atmospheric conditions such as a change in water vapor or dust content in the
atmosphere or a change in the ozone layer in the upper atmosphere also affects the
zero point. It is known that the change in water vapor content affects the extinction at the
longer wavelength (mostly $I$ and near-IR), while changes in the thickness of the ozone layer affects
the extinction at the short wavelength ($U$ and $B$) (see SB00 or Sung et al 2001).

Figure 4 shows a typical case of time variation obtained on 2004, August 15. We found that in many
cases the time variation of the photometric zero-points at MAO started at evening twilight and ended
around midnight ($UT = 20 \sim 21 ^h$). We made a note in the 8th column of Table 1 if there was
a noticeable amount of time variation during the night.

\begin{figure}[!t]
\epsfxsize=9cm \epsfbox{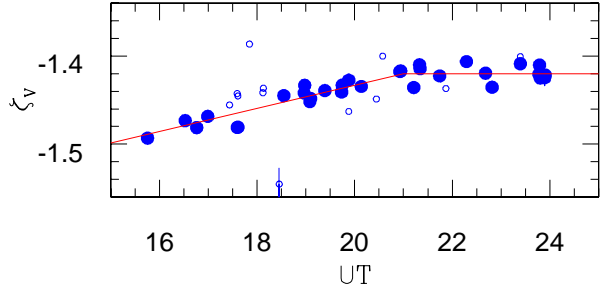} \caption{The variation of the photometric zero point in the $V$ filter
against UT on 2004, August 15. The open circles represent variable star candidates. The photometric zero
point slightly increased between $UT = 14^h$ and $21^h$ with no variation after $UT = 21^h$.}
\end{figure}

\subsection{Standard Transformation}
We transformed the data to the SAAO system using the relations above, and present the results
in Table 3. The data for variable star candidates in Table 5 are not listed in the table.
We also plot the residuals relative to the SAAO system in Figure 5 for the 2k CCD and
in Figure 6 for the 4k CCD. The number at the bottom of each panel represents the weighted
mean value and standard deviation of the residuals ($weight =$ the number of observations).
The mean value indicates the difference of zero points of our results relative to the SAAO
system, while the standard deviation represents the precision (or reproducibility) of our
photometry. In general, the differences in photometric zero point relative to the SAAO system are
less than 1 $mmag$. But in $U-B$ (and $V-I$ from the 2k CCD) the difference is slightly larger ( 3 -- 3.6 $mmag$).
A somewhat larger difference in the $V-I$ zeropoint of the 2k CCD is probably
due to the lack of red standard stars.
The slope of the $I$ transformation is strongly affected by the color where the slope changes. Unfortunately only
one star with $V-I > $ 1.6 was secured and the 2k CCD was replaced by a new 4k CCD in 2006, June.

\begin{table*}
\begin{center}
\scriptsize
\bf{\sc  Table 2.}\\
\sc{Transformation coefficients} \\
\begin{tabular}{c|c|cc|cc}
\\ \hline \hline
          &             & \multicolumn{2}{c}{SITe 2k CCD} & \multicolumn{2}{c}{Fairchild 4k CCD} \\
 Filter   & color & $\eta$  & color range  & $\eta$ &  color range \\
\hline
          &          &                    &                  &                     &     \\

          & $V-I$    & $+0.054 \pm 0.002$ &  $V-I \leq 1.6 $ & $+0.041 \pm 0.001$  & All \\
   $I$    &          & $0.000$            &  $V-I > 1.6 $    &                     &     \\
          & $R-I$    & $+0.113 \pm 0.018$ &  $R-I \leq 0.75$ & $+0.086 \pm 0.002$  & All \\
          &          & $0.000$            &  $R-I > 0.75$    &                     &     \\ \hline
   $R$    & $R-I$    &         -          &        -         & $-0.065 \pm 0.003$  & All \\ \hline
          & $B-V$    & $-0.043 \pm 0.002$ &  All             & $-0.061 \pm 0.001$  & $B-V \leq 1.6 $ \\
          &          &                    &                  &      0.000          & $B-V > 1.6 $ \\
   $V$    & $V-I$    & $-0.044 \pm 0.002$ &  $V-I \leq 1.5$  & $-0.063 \pm 0.001$  & $V-I \leq 1.6$ \\
          &          & $0.000$            &  $V-I > 1.5$     & $0.000           $  & $V-I > 1.6$ \\ \hline
   $B$    & $B-V$    & $+0.053 \pm 0.003$ &  $B-V \leq 1.2$  & $+0.025 \pm 0.001$  & All \\
          &          & $0.000$            &  $B-V > 1.2$     &                     &     \\ \hline
          & $U-B$    & $+0.103 \pm 0.005$ &  $U-B \leq 0.0$  & $f[(B-V)_0]      $  & Early stars \\
   $U$    &          & $+0.010 \pm 0.003$ &  $U-B > 0.0   $  & $+0.057 \pm 0.002$  & $U-B \leq 1.0$ \\
          &          &                    &                  & $+0.116 \pm 0.005$  & $U-B > 1.0$ \\
 \hline\\
\end{tabular}
\end{center}
\end{table*}

\begin{figure}[!]
\epsfxsize=9cm \epsfbox{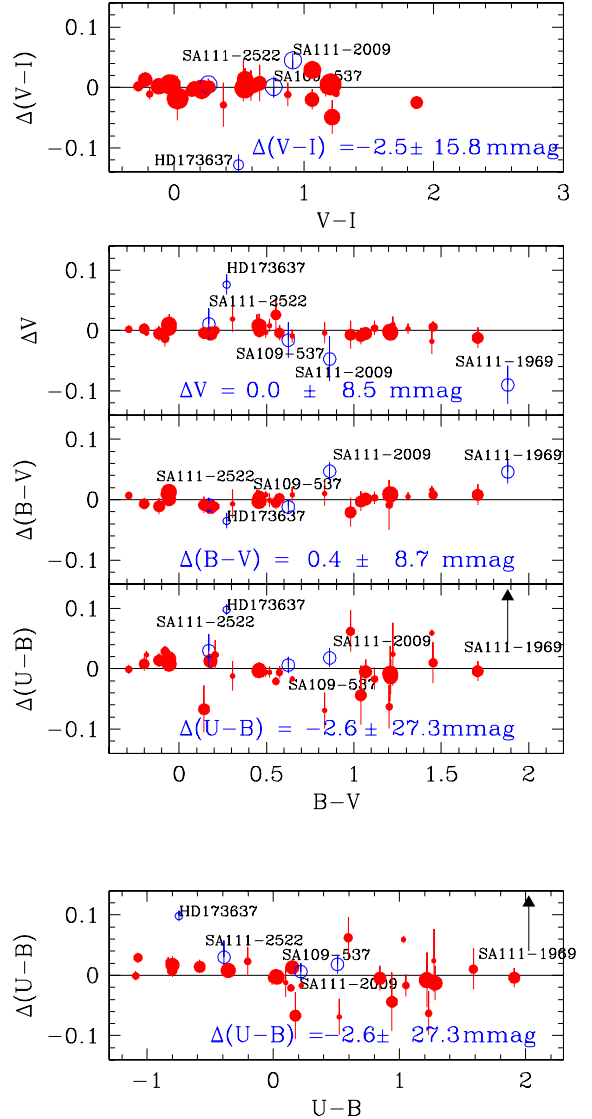}
\caption{Residuals of the SITe 2k CCD data
relative to the SAAO system against an appropriate color index. The numerical value at
the bottom of each panel represents the weighted mean value and standard deviation
of the residuals.
The size of circles is proportional to the numbers of our observations. The
open circle denotes variable star candidates. The meaning of $\Delta$ is in the sense SAAO data minus the
transformed 2k data.}
\end{figure}

\begin{figure}[!]
\epsfxsize=9cm \epsfbox{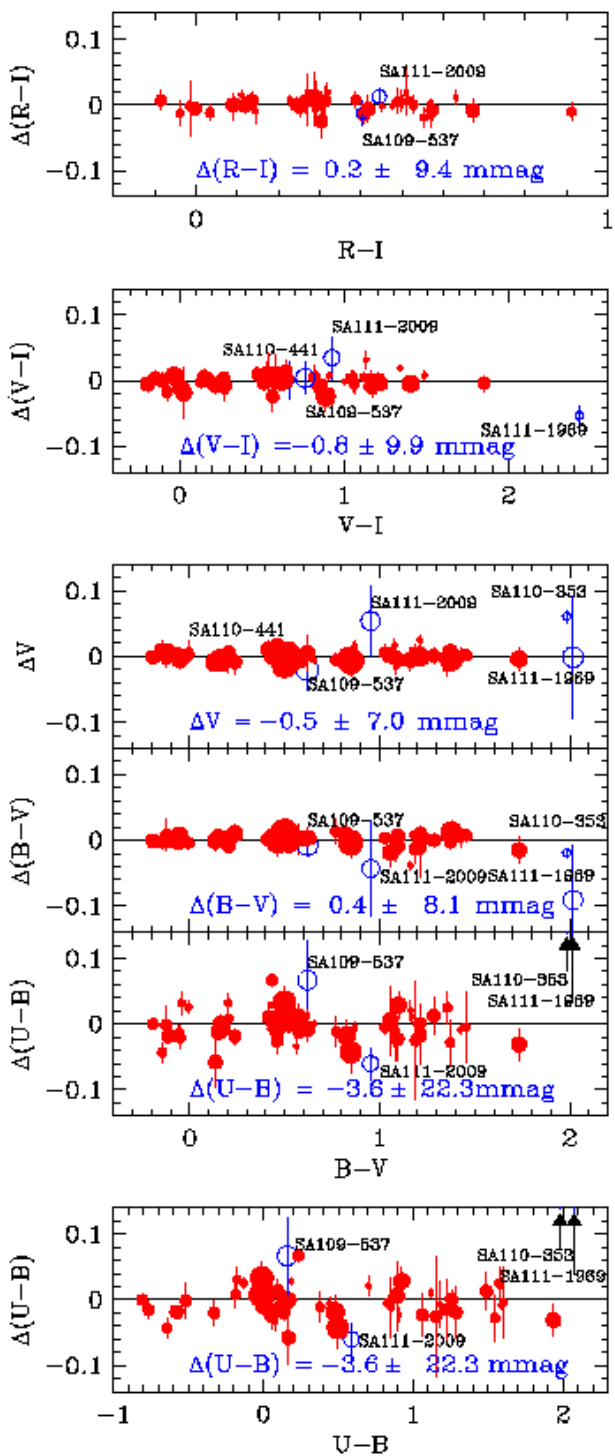} \caption{Residuals of the Fairchild 486 CCD data
relative to the SAAO system. $\Delta$ represents the SAAO data minus the
transformed 4K data. The other symbols are the same as Figure 5.}
\end{figure}

\begin{figure}[!t]
\epsfxsize=9cm \epsfbox{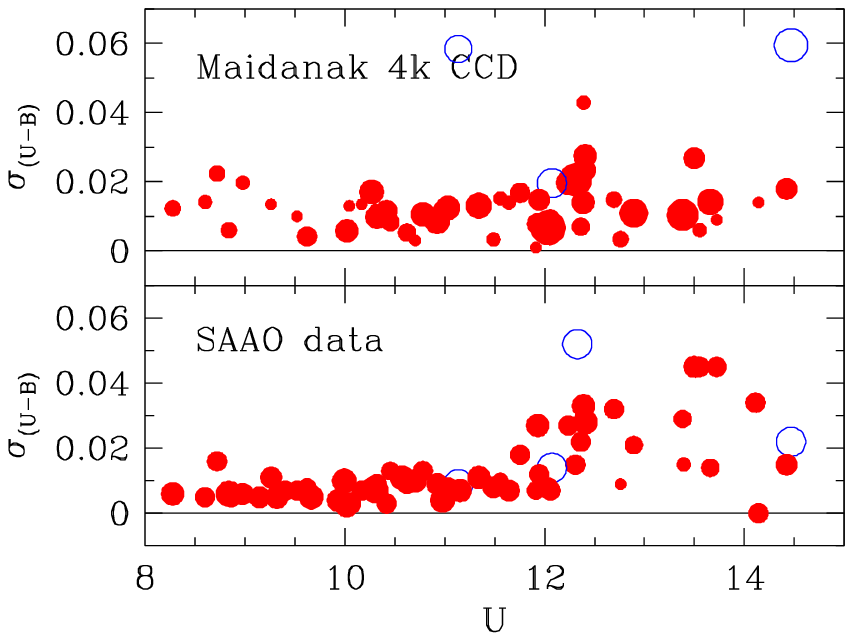} \caption{Comparison of the photometric
errors in $U-B$ of the 4K CCD data and SAAO data against $U$ magnitude. The size of circles
is proportional to the number of observations. The open circles represent
variable candidates. The standard deviation of the 4K data is in general smaller than 0.02 $mag$,
while that of the SAAO data shows a drastic increase in scatter for $U >$ 12 $mag$.}
\end{figure}

For the $U-B$ color, both 2K (CCD data) and 4K CCD data show a similar size zero point offset and
a larger scatter. The error in the bottom panel of Figure 2 is mainly due to the error in the
M91 photometry, not in ours. At SAAO the $50cm$ telescope was used for the photometry of
standard stars and the quantum efficiency of the photomultiplier tubes was much lower than that of our CCD.
As shown in Figure 7, the errors in $U-B$ of the SAAO data increase rapidly for $U\geq12$, while those in
our data are nearly flat down to $U\approx14$. A somewhat larger offset in zero point and a
larger scatter between our data and the SAAO data is inevitable.

\begin{table*}
\begin{center}
\scriptsize
\bf{\sc  Table 3.}\\
\sc{Photometric data: standard stars} \\
\begin{tabular}{cccccccccccccccccccc}
\\ \hline \hline
 Star      &   $V$  &  $R-I$  & $V-I$  &  $B-V$ &  $U-B$ &  $\epsilon_V$ & $\epsilon_{R-I}$ & $\epsilon_{V-I}$ & $\epsilon_{B-V}$ & $\epsilon_{U-B}$
&\multicolumn{5}{c}{$n_{obs}$} \\
 \hline\\
 BD+5 2468 &   9.359  &   -0.039  &  -0.077  &  -0.122  &  -0.518  &   0.006  &   0.001  &   0.004  &   0.023  &   0.022  &   4  &  2  &  4 &   4 &   4 \\
 BD-11 162 &  11.181  &      -    &   0.135  &  -0.073  &  -1.072  &   0.003  &     -    &   0.001  &   0.005  &   0.007  &  7   & 0   & 7  &  7  &   3 \\
 BD-15 115 &  10.867  &      -    &  -0.223  &  -0.200  &  -0.801  &   0.002  &     -    &   0.007  &   0.005  &   0.009  &  4   &  0  & 4  &  4  &   4 \\
 BD-2 524  &  10.312  &      -    &  -0.104  &  -0.112  &  -0.579  &   0.003  &     -    &   0.004  &   0.006  &   0.004  &   6  &  0  &  6 &   6 &   6 \\
 HD 149382 &   8.948  &      -    &  -0.276  &  -0.288  &  -1.090  &   0.005  &     -    &   0.003  &   0.004  &   0.006  &  2   &  0  & 2  &  2  &   2 \\
 HD 160233 &   9.098  &   -0.016  &  -0.030  &  -0.058  &  -0.789  &   0.001  &   0.004  &   0.001  &   0.001  &   0.015  &  18  & 2   & 18 &  17 &  11 \\
 HD 209684 &   9.845  &      -    &  -0.191  &  -0.190  &  -0.828  &   0.001  &     -    &   0.002  &   0.003  &   0.009  &  7   & 0   & 7  &  7  &   5 \\
 HD 216135 &  10.119  &      -    &  -0.118  &  -0.114  &  -0.583  &   0.007  &     -    &   0.008  &   0.008  &   0.009  &  5   &  0  & 5  &  5  &   5 \\
 P1633+099 &  14.360  &      -    &  -0.177  &  -0.202  &  -1.010  &   0.004  &     -    &   0.012  &   0.005  &   0.005  &  1   &  0  & 1  &  1  &   1 \\
P1633+099A &  15.237  &      -    &   1.007  &   0.870  &     -    &   0.008  &     -    &   0.012  &   0.014  &   0.000  &  1   &  0  & 1  &  1  &   0 \\
P1633+099B &  12.949  &      -    &   1.069  &   1.078  &   0.998  &   0.002  &     -    &   0.003  &   0.004  &   0.009  &  1   &  0  & 1  &  1  &   1 \\
P1633+099C &  13.203  &      -    &   1.112  &   1.135  &   1.119  &   0.002  &     -    &   0.003  &   0.004  &   0.012  &  1   &  0  & 1  &  1  &   1 \\
P1633+099D &  13.662  &      -    &   0.638  &   0.526  &  -0.031  &   0.003  &     -    &   0.004  &   0.004  &   0.003  &  1   &  0  & 1  &  1  &   1 \\
 SA92-263  &  11.791  &    0.510  &   1.064  &   1.084  &   0.851  &   0.002  &   0.001  &   0.002  &   0.008  &   0.009  &   1  &  1  &  1 &   1 &   2 \\
 SA92-330  &  15.041  &      -    &   0.728  &   0.578  &  -0.055  &   0.020  &     -    &   0.028  &   0.025  &   0.019  &  1   &  0  & 1  &  1  &   1 \\
 SA92-335  &  12.534  &      -    &   0.727  &   0.685  &   0.245  &   0.006  &     -    &   0.003  &   0.008  &   0.005  &  3   &  0  & 3  &  3  &   3 \\
 SA92-339  &  15.556  &      -    &   0.631  &   0.514  &  -0.197  &   0.034  &     -    &   0.047  &   0.041  &   0.026  &  1   &  0  & 1  &  1  &   1 \\
 SA92-342  &  11.616  &    0.272  &   0.533  &   0.446  &  -0.009  &   0.006  &   0.005  &   0.001  &   0.002  &   0.005  &  9   & 3   & 6  &  8  &  11 \\
 SA93-241  &   9.393  &    0.452  &   0.909  &   0.868  &   0.443  &   0.001  &   0.001  &   0.001  &   0.001  &   0.003  &   1  &  1  &  1 &   1 &   1 \\
 SA93-317  &  11.551  &    0.293  &   0.588  &   0.498  &  -0.019  &   0.005  &   0.010  &   0.003  &   0.002  &   0.006  &  20  & 7   & 12 &  18 &  15 \\
 SA93-326  &   9.572  &    0.259  &   0.521  &   0.454  &  -0.004  &   0.003  &   0.010  &   0.002  &   0.004  &   0.002  &  12  & 5   & 7  &  11 &   8 \\
 SA93-332  &   9.789  &    0.288  &   0.598  &   0.519  &   0.013  &   0.000  &   0.003  &   0.007  &   0.002  &   0.002  &  12  & 6   & 8  &  11 &   9 \\
 SA93-333  &  12.027  &    0.418  &   0.881  &   0.844  &   0.502  &   0.001  &   0.004  &   0.006  &   0.005  &   0.010  &  21  & 6   & 12 &  19 &  15 \\
 SA93-424  &  11.636  &    0.499  &   1.060  &   1.098  &   0.924  &   0.006  &   0.004  &   0.006  &   0.011  &   0.014  &   5  &  2  &  4 &   5 &   8 \\
 SA94-242  &  11.729  &      -    &   0.380  &   0.306  &   0.099  &   0.001  &     -    &   0.003  &   0.001  &   0.001  &  1   &  0  & 1  &  1  &   1 \\
 SA94-251  &  11.214  &      -    &   1.248  &   1.224  &   1.272  &   0.001  &     -    &   0.001  &   0.001  &   0.002  &  1   &  0  & 1  &  1  &   1 \\
 SA94-305  &   8.902  &      -    &     -    &   1.424  &   1.587  &   0.001  &     -    &     -    &   0.001  &   0.001  &   1  &  0  &  0 &   1 &   1 \\
 SA94-308  &   8.753  &    0.289  &   0.573  &   0.500  &   0.008  &   0.001  &   0.001  &   0.001  &   0.001  &   0.001  &   1  &  1  &  1 &   1 &   1 \\
 SA95-132  &  12.094  &    0.305  &   0.565  &   0.433  &   0.232  &   0.006  &   0.006  &   0.006  &   0.009  &   0.003  &   7  &  5  &  5 &   6 &   4 \\
  SA95-74  &  11.536  &    0.574  &   1.169  &   1.161  &   0.703  &   0.005  &   0.007  &   0.006  &   0.002  &   0.002  &   7  &  5  &  5 &   1 &   1 \\
 SA96-180  &   8.928  &      -    &     -    &   1.075  &     -    &   0.001  &     -    &     -    &   0.001  &     -    &   1  &  0  &  0 &   1 &   0 \\
  SA96-36  &  10.604  &    0.134  &   0.275  &   0.238  &   0.136  &   0.010  &   0.005  &   0.005  &   0.011  &   0.012  &   6  &  6  &  6 &   8 &   5 \\
 SA96-393  &   9.659  &    0.319  &   0.674  &   0.603  &   0.044  &   0.003  &   0.001  &   0.011  &   0.001  &   0.012  &   2  &  1  &  2 &   4 &   2 \\
 SA96-406  &   9.300  &    0.110  &   0.246  &   0.206  &   0.156  &   0.001  &   0.001  &   0.001  &   0.001  &   0.001  &   1  &  1  &  1 &   1 &   1 \\
 SA97-284  &  10.790  &      -    &   1.488  &     -    &     -    &   0.001  &     -    &   0.001  &     -    &     -    &   1  &  0  &  1 &   0 &   0 \\
 SA97-346  &   9.270  &      -    &     -    &   0.617  &   0.105  &   0.000  &     -    &     -    &   0.001  &   0.001  &   2  &  0  &  0 &   2 &   1 \\
 SA97-351  &   9.797  &    0.144  &   0.274  &   0.203  &   0.038  &   0.006  &   0.001  &   0.001  &   0.012  &   0.013  &   4  &  1  &  2 &   4 &   2 \\
 SA98-193  &  10.039  &    0.553  &   1.148  &   1.190  &   1.152  &   0.007  &   0.008  &   0.004  &   0.010  &   0.014  &   5  &  2  &  3 &   5 &   4 \\
 SA98-653  &   9.540  &    0.034  &   0.033  &  -0.007  &  -0.135  &   0.004  &   0.007  &   0.002  &   0.008  &   0.005  &   4  &  3  &  3 &   4 &   2 \\
 SA98-978  &  10.584  &    0.315  &   0.648  &   0.619  &   0.139  &   0.007  &   0.008  &   0.006  &   0.010  &   0.013  &   7  &  4  &  5 &   7 &   7 \\
 SA99-296  &   8.459  &    0.517  &   1.115  &   1.211  &   1.257  &   0.006  &   0.001  &   0.004  &   0.014  &   0.009  &   7  &  1  &  2 &   6 &   5 \\
 SA99-408  &   9.816  &    0.243  &   0.480  &   0.414  &   0.037  &   0.006  &   0.002  &   0.006  &   0.008  &   0.017  &   7  &  5  &  5 &   6 &   5 \\
 SA99-418  &   9.471  &   -0.002  &  -0.009  &  -0.044  &  -0.176  &   0.002  &   0.005  &   0.006  &   0.005  &   0.014  &   7  &  5  &  5 &   5 &   2 \\
 SA99-438  &   9.397  &   -0.085  &  -0.149  &  -0.145  &  -0.645  &   0.012  &   0.007  &   0.004  &   0.009  &   0.014  &   6  &  4  &  4 &   4 &   3 \\
 SA100-162 &   9.156  &    0.570  &   1.200  &   1.284  &   1.490  &   0.003  &   0.015  &   0.002  &   0.007  &   0.008  &   5  &  1  &  2 &   5 &   5 \\
 SA100-241 &  10.149  &    0.091  &   0.163  &   0.153  &   0.116  &   0.006  &   0.006  &   0.004  &   0.007  &   0.012  &   9  &  7  &  7 &   8 &   7 \\
 SA100-280 &  11.816  &    0.288  &   0.579  &   0.500  &  -0.009  &   0.006  &   0.013  &   0.010  &   0.007  &   0.021  &  19  &  7  & 10 &  19 &  16 \\
 SA101-281 &  11.590  &    0.407  &   0.854  &   0.826  &   0.480  &   0.007  &   0.008  &   0.005  &   0.015  &   0.011  &   7  &  6  &  7 &   7 &  10 \\
 SA101-282 &  10.010  &    0.252  &   0.509  &   0.435  &   0.008  &   0.005  &   0.008  &   0.009  &   0.006  &   0.008  &   4  &  4  &  4 &   4 &   5 \\
 SA102-466 &   9.255  &    0.502  &   1.054  &     -    &     -    &   0.005  &   0.007  &   0.005  &     -    &     -    &   2  &  2  &  2 &   0 &   0 \\
 SA102-472 &   8.754  &    0.478  &   1.001  &   1.027  &   0.835  &   0.010  &   0.001  &   0.001  &   0.002  &   0.005  &   4  &  1  &  1 &   4 &   3 \\
 SA102-620 &  10.080  &    0.528  &   1.160  &   1.095  &   1.063  &   0.007  &   0.002  &   0.003  &   0.004  &   0.020  &   6  &  3  &  3 &   6 &   5 \\
 SA102-625 &   8.889  &    0.300  &   0.609  &   0.560  &   0.076  &   0.003  &   0.001  &   0.001  &   0.008  &   0.010  &   4  &  2  &  2 &   4 &   1 \\
 SA103-462 &  10.119  &    0.292  &   0.617  &   0.568  &   0.095  &   0.003  &   0.007  &   0.006  &   0.021  &   0.011  &   6  &  5  &  4 &   6 &   5 \\
 SA103-483 &   8.342  &    0.227  &     -    &   0.421  &   0.099  &   0.001  &   0.001  &     -    &   0.001  &   0.001  &   1  &  1  &  0 &   1 &   1 \\
 SA104-337 &  11.216  &    0.388  &   0.820  &   0.771  &   0.376  &   0.009  &   0.006  &   0.015  &   0.017  &   0.007  &   4  &  3  &  4 &   4 &   3 \\
 SA104-461 &   9.705  &    0.290  &   0.567  &   0.488  &  -0.025  &   0.015  &   0.003  &   0.009  &   0.015  &   0.014  &   2  &  2  &  2 &   2 &   2 \\
 SA105-205 &   8.820  &      -    &     -    &   1.356  &   1.578  &   0.012  &     -    &     -    &   0.004  &   0.017  &   3  &  0  &  0 &   3 &   3 \\
 SA106-700 &   9.783  &    0.631  &   1.340  &   1.371  &   1.546  &   0.005  &   0.010  &   0.001  &   0.010  &   0.015  &   6  &  1  &  1 &   5 &   3 \\
 SA107-592 &  11.856  &      -    &     -    &   1.323  &     -    &   0.001  &     -    &     -    &   0.009  &     -    &   1  &  0  &  0 &   1 &   0 \\
 SA108-475 &  11.319  &    0.672  &   1.403  &   1.377  &   1.437  &   0.012  &   0.006  &   0.008  &   0.008  &   0.014  &  15  &  6  &  9 &  14 &   2 \\
 SA108-551 &  10.706  &    0.119  &   0.214  &   0.177  &   0.153  &   0.001  &   0.006  &   0.002  &   0.005  &   0.004  &  19  & 6   & 14 &  16 &  15 \\
 SA108-827 &   7.957  &      -    &     -    &   1.310  &     -    &   0.001  &     -    &     -    &   0.001  &     -    &  1   &  0  & 0  &  1  &   0 \\
 SA109-231 &   9.331  &      -    &     -    &   1.454  &   1.586  &   0.002  &     -    &     -    &   0.000  &   0.002  &  7   & 0   & 0  &  7  &   5 \\
 SA109-243 &  11.954  &      -    &     -    &   0.590  &   0.312  &   0.004  &     -    &     -    &   0.005  &   0.003  &  1   &  0  & 0  &  1  &   1 \\
 SA109-255 &  11.142  &      -    &   1.009  &   0.939  &   0.632  &   0.002  &     -    &   0.003  &   0.003  &   0.003  &  1   &  0  & 1  &  1  &   1 \\
 SA110-280 &  13.000  &      -    &     -    &   2.130  &     -    &   0.001  &     -    &     -    &   0.003  &     -    &   2  &  0  &  0 &   1 &   0 \\
 SA110-355 &  11.953  &      -    &     -    &   1.049  &     -    &   0.001  &     -    &     -    &   0.001  &     -    &   2  &  0  &  0 &   2 &   0 \\
 SA110-450 &  11.574  &      -    &   1.217  &   0.981  &   0.594  &   0.011  &     -    &   0.009  &   0.014  &   0.002  &  5   &  0  & 5  &  5  &   3 \\
SA111-2522 &   9.688  &      -    &   0.264  &   0.170  &  -0.390  &   0.011  &     -    &   0.008  &   0.009  &   0.023  &  3   &  0  & 3  &  3  &   3 \\
 SA111-717 &   8.530  &      -    &   0.469  &   0.424  &   0.186  &   0.001  &     -    &   0.006  &   0.008  &   0.001  &   2  &  0  &  2 &   2 &   1 \\
 SA111-773 &   8.968  &    0.139  &   0.269  &   0.205  &  -0.193  &   0.003  &   0.008  &   0.001  &   0.001  &   0.007  &  8   & 2   & 8  &  8  &   5 \\
 SA111-775 &  10.749  &    0.912  &   1.864  &   1.710  &   1.912  &   0.004  &   0.006  &   0.009  &   0.006  &   0.008  &  13  & 3   & 8  &  12 &  12 \\
 SA112-223 &  11.429  &    0.266  &   0.544  &   0.457  &   0.032  &   0.002  &   0.000  &   0.003  &   0.001  &   0.005  &  15  & 4   & 12 &  14 &  13 \\
 SA112-250 &  12.108  &      -    &   0.632  &   0.536  &  -0.005  &   0.002  &     -    &   0.004  &   0.004  &   0.004  &  8   &  0  & 8  &  8  &   8 \\
 SA112-275 &   9.914  &    0.570  &   1.210  &   1.210  &   1.281  &   0.002  &   0.001  &   0.007  &   0.000  &   0.001  &  16  & 2   & 4  &  15 &  13 \\
 SA112-805 &  12.095  &    0.091  &   0.153  &   0.141  &   0.170  &   0.002  &   0.001  &   0.003  &   0.003  &   0.005  &  16  & 4   & 11 &  13 &  11 \\
 SA112-810 &  10.576  &      -    &   1.095  &   1.148  &   1.142  &   0.001  &     -    &   0.001  &   0.002  &   0.002  &  1   &  0  & 1  &  1  &   1 \\
 SA112-822 &  11.553  &    0.491  &   1.042  &   1.044  &   0.921  &   0.004  &   0.009  &   0.007  &   0.007  &   0.024  &  13  & 3   & 8  &  13 &  12 \\
 SA113-259 &  11.752  &      -    &   1.161  &   1.211  &   1.186  &   0.006  &     -    &   0.013  &   0.003  &   0.023  &  3   & 0   & 3  &  3  &   3 \\
 SA113-269 &   9.489  &      -    &     -    &   1.119  &   1.051  &   0.003  &     -    &     -    &   0.009  &   0.008  &  2   &  0  & 0  &  2  &   2 \\
 SA113-274 &   8.829  &      -    &     -    &   0.483  &   0.007  &   0.001  &     -    &     -    &   0.001  &   0.003  &   1  &  0  &  0 &   1 &   1 \\
 SA113-276 &   9.080  &      -    &     -    &   0.653  &   0.220  &   0.000  &     -    &     -    &   0.000  &   0.006  &  2   & 0   & 0  &  2  &   2 \\
 SA113-466 &  10.012  &      -    &   0.546  &   0.459  &   0.016  &   0.005  &     -    &   0.006  &   0.005  &   0.002  &  5   &  0  & 5  &  5  &   5 \\
 SA113-475 &  10.315  &      -    &   1.065  &   1.067  &   0.845  &   0.005  &     -    &   0.011  &   0.004  &   0.006  &  6   &  0  & 6  &  6  &   6 \\
 SA114-670 &  11.118  &    0.578  &   1.212  &   1.209  &   1.219  &   0.000  &   0.001  &   0.004  &   0.003  &   0.003  &  19  & 1   & 17 &  17 &  13 \\
 SA114-750 &  11.924  &   -0.011  &   0.026  &  -0.056  &  -0.355  &   0.005  &   0.001  &   0.000  &   0.002  &   0.007  &  19  & 2   & 17 &  17 &  12 \\
 SA114-755 &  10.914  &    0.309  &   0.618  &   0.573  &  -0.004  &   0.000  &   0.002  &   0.003  &   0.000  &   0.002  &  14  & 1   & 12 &  12 &   6 \\
 SA115-349 &   8.585  &      -    &     -    &   1.077  &   0.909  &   0.001  &     -    &     -    &   0.001  &   0.001  &   1  &  0  &  0 &   1 &   1 \\
 SA115-420 &  11.177  &    0.275  &   0.570  &   0.474  &  -0.009  &   0.010  &   0.005  &   0.011  &   0.012  &   0.014  &   2  &  2  &  2 &   2 &   2 \\
 SA115-427 &   8.865  &      -    &     -    &   1.170  &   1.120  &   0.001  &     -    &     -    &   0.001  &   0.001  &   1  &  0  &  0 &   1 &   1 \\
\\
\hline
\end{tabular}
\end{center}
\end{table*}

\section{Comparisons with Landolt's Photometry}
Small systematic differences between the SAAO system and Landolt's version is well known (M91),
and is confirmed by SB00. We compare our data with L83 and L92 in Figures 8 and 9 for the 2k CCD and
in Figures 10 and 11 for the 4k CCD. The difference ($\Delta \equiv $Landolt - our data) is shown
plotted against an appropriate color. In addition, we superimpose the mean line of the systematic
difference between the SAAO system and Landolt system from M91. The differences between our data
and Landolt's photometry follows well the trend obtained by M91, and therefore we could confirm
again the differences between the two systems. This fact indirectly implies that we have reproduced the
SAAO system at MAO. We will make some notes on each color in this section.

\begin{figure}[!]
\epsfxsize=9cm \epsfbox{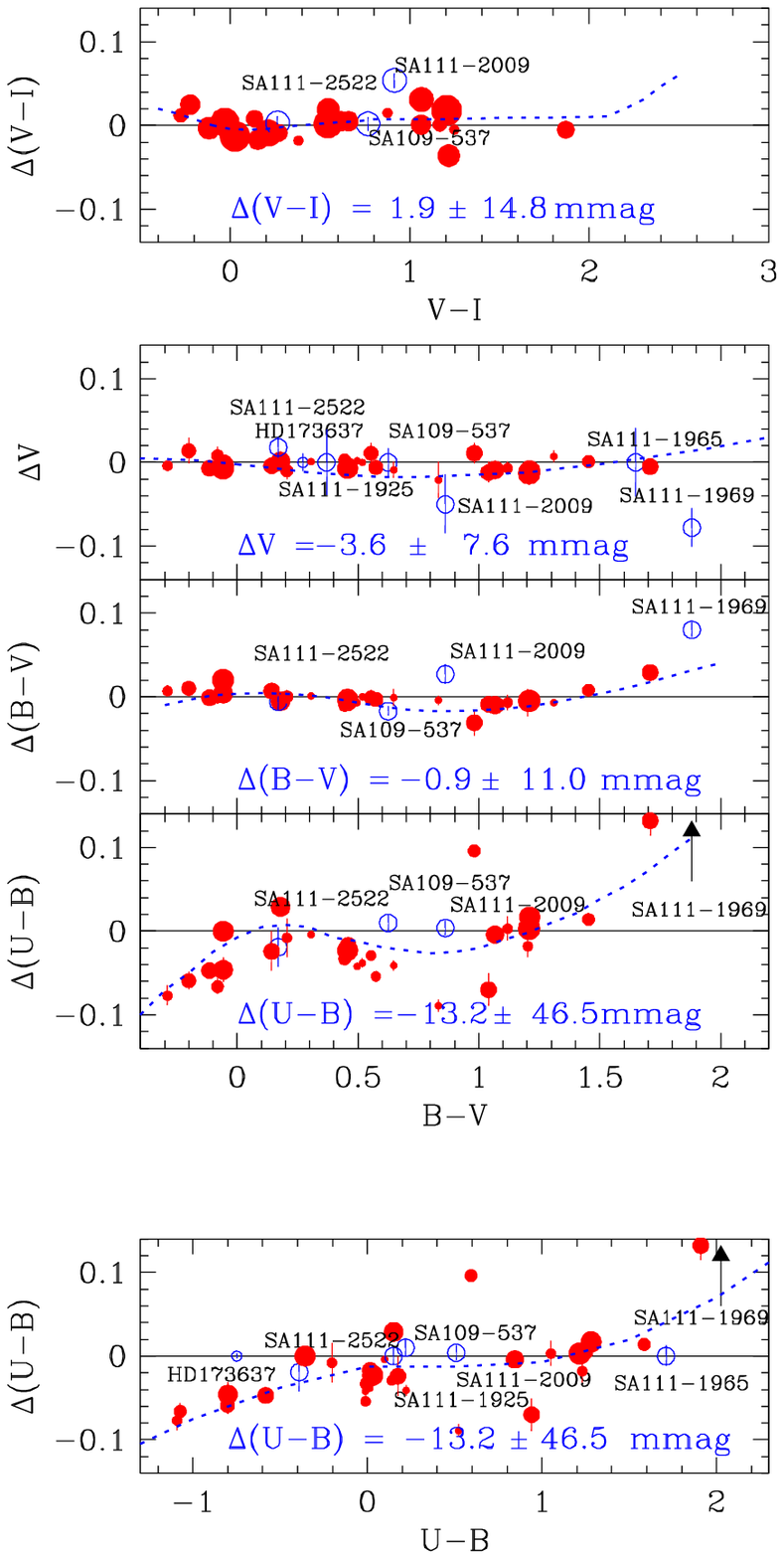} \caption{Differences in the SITe 2k CCD data
relative to Landolt (1983) against an appropriate color index. $\Delta$ represents Landolt (1983)
minus the transformed 2k data. The dotted line represents the mean difference
between the SAAO system and the Landolt system from Menzies et al. (1991).
The other symbols are the same as Figure 5.}
\end{figure}

\begin{figure}[!]
\epsfxsize=9cm \epsfbox{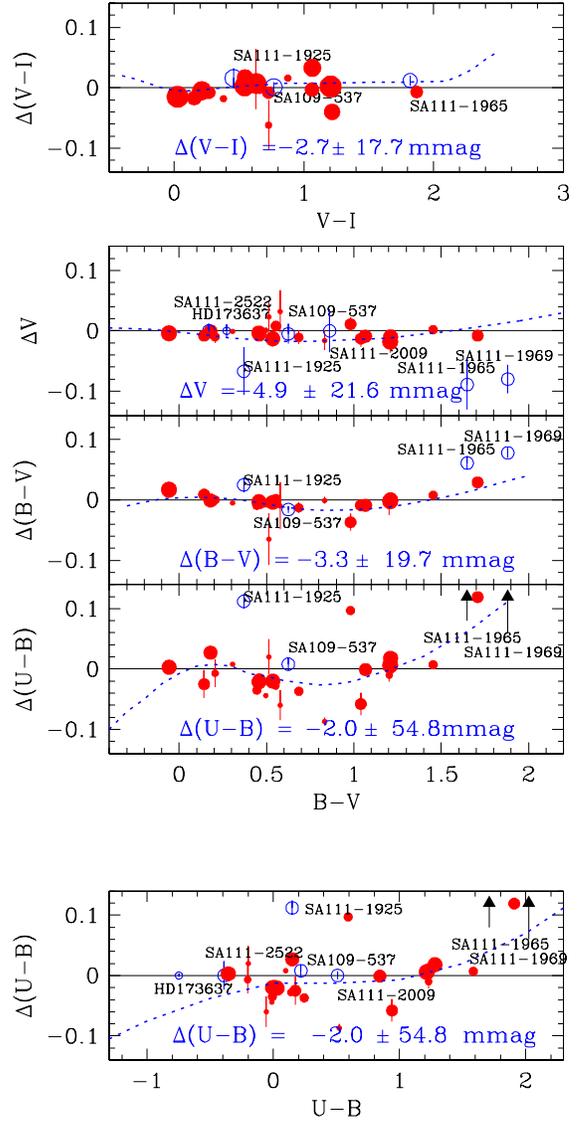} \caption{Differences in the SITe 2k CCD data
relative to Landolt (1992). $\Delta$ represents Landolt (1992)
minus the transformed 2k data. The other symbols are the same as Figure 8.}
\end{figure}

\begin{figure}[!]
\epsfxsize=9cm \epsfbox{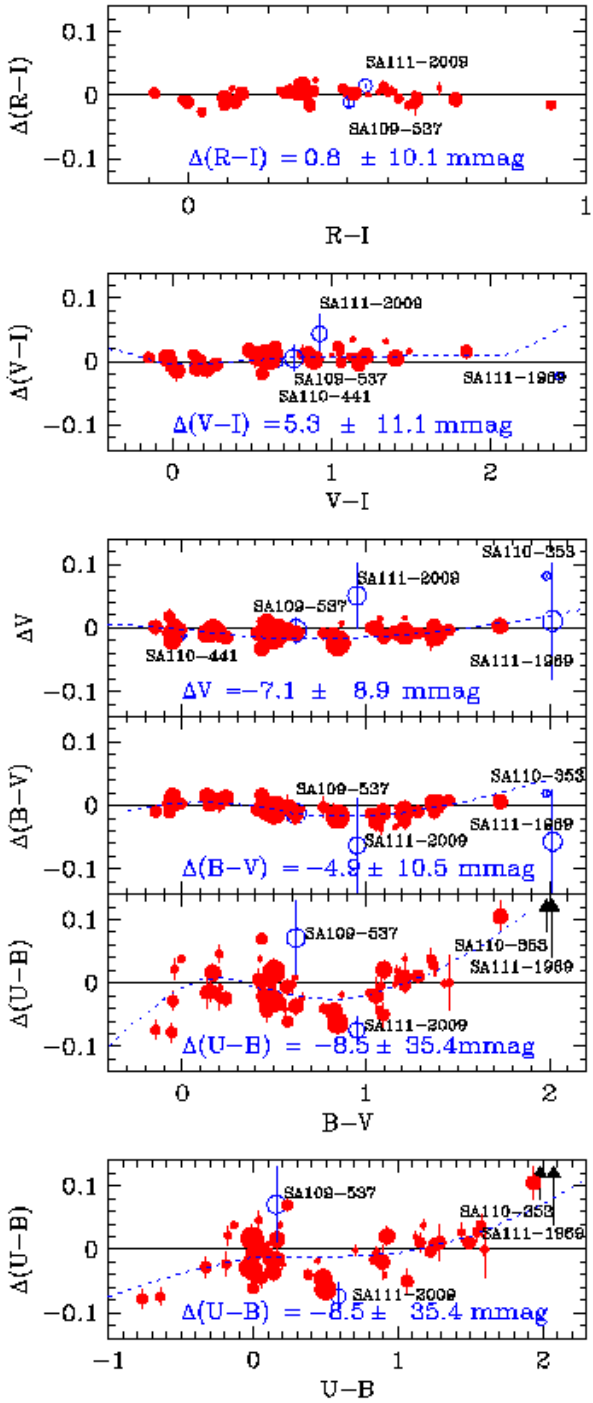} \caption{Differences in the Fairchild 486 CCD data
relative to Landolt (1983). $\Delta$ represents Landolt (1983)
minus the transformed 4k data. The other symbols are the same as Figure 8.}
\end{figure}

\begin{figure}[!]
\epsfxsize=9cm \epsfbox{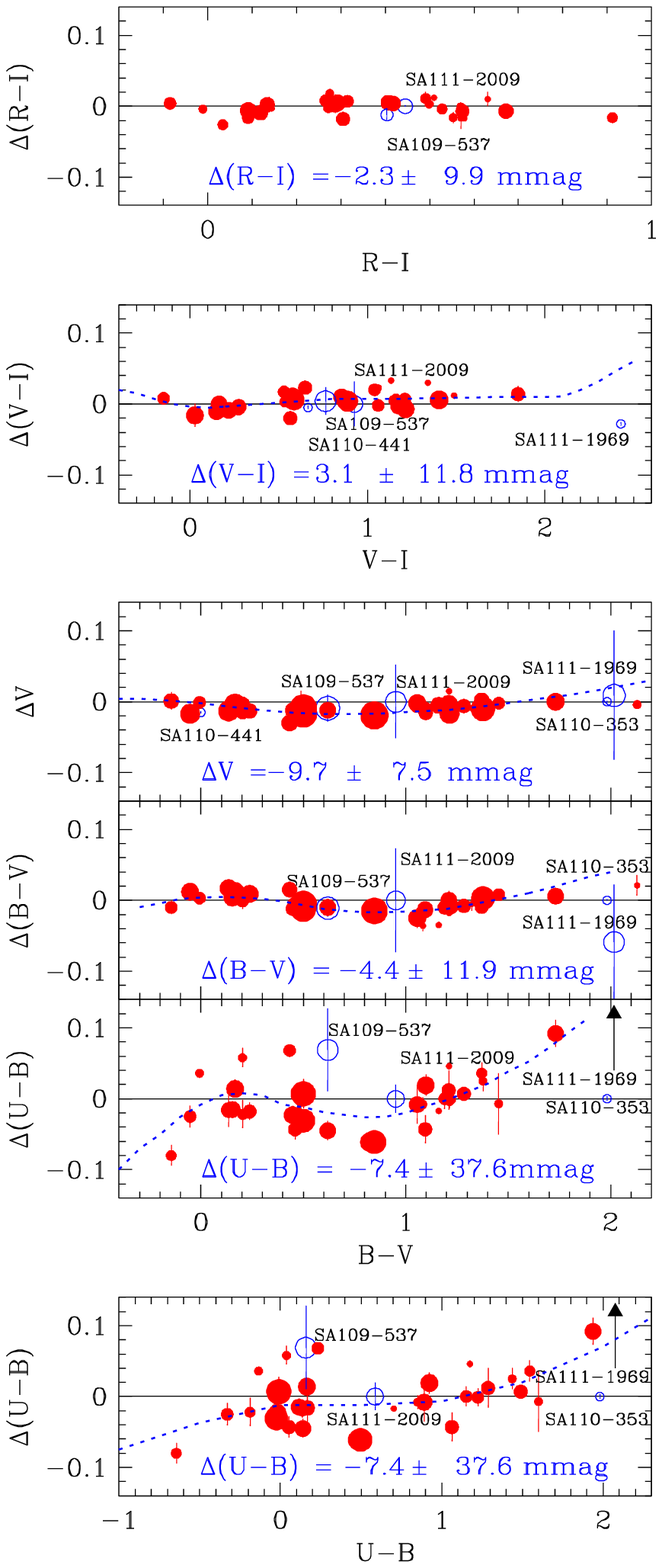} \caption{Difference in the Fairchild 486 CCD data
relative to Landolt (1992) for the stars in Menzies et al. (1991) or Kilkenny et al. (1998).
$\Delta$ represents Landolt (1992) minus the transformed
4k data. The other symbols are the same as Figure 8.}
\end{figure}

\subsection{$R-I$}

M91 did not make any comparison with L83, and we did not observe many standard stars
in $R$ with the 2k CCD at MAO, therefore we derived the transformation relation only
for the 4k CCD. There is no remarkable difference between our data and L83 or L92. The two
data sets match well within 2$mmag$ with a standard deviation of 0.01 $mag$

\subsection{$V-I$}

The difference between the SAAO system and Landolt's data in $V-I$ is not large (less than 0.01 $mag$)
for $V-I=$ 0 -- 2, but the Landolt data are redder for bluer or redder stars. We could not confirm
the difference for red stars ($V-I > 2$) due to the lack of very red stars in our data. But
our 2k data show clearly that L83 data are redder than ours for blue stars, which is the same trend found by M91.

\subsection{$V$}

M91 found that Landolt $V$ data are slightly brighter (about 5 $mmag$) for $B-V \geq 0.5$ and has a color
dependency for bluer stars ($B-V < 0.5$). We could not confirm the color dependency for blue stars
from the 2k CCD data. But we could confirm the zero point difference of about 5 $mmag$ (3.6--9.7 $mmag$).
The distribution of $\Delta V$ of the 4k data well follows the trend obtained by M91.

\subsection{$B-V$}

Differences between the two standard systems in $B-V$ shows a sinusoidal-like curve against $B-V$.
As the SAAO system well reproduced the original Johnson's $B-V$ (Cousins 1984), this
implies that Landolt's $B$ filter is more affected by the size of the Balmer jump or the strength of the hydrogen lines.
The average zero point
difference between our data and L83 or L92 is small, i.e. less than 5 $mmag$ with a standard deviation
of about 0.01 $mag$, but the differences against $B-V$ for both L83 and L92 well follow the trend
found by M91. This implies that L92 has the same difference relative to the SAAO system.

\subsection{$U-B$}

Although systematic differences exist between the SAAO system and the Landolt system, the
differences in most color indices are quite small ($\leq 0.02$ $mag$) for normal stars ($B-V\leq1.5$ or
$V-I < 2$). But the difference in $U-B$ is too large to be ignored. Although the $U-B$ scale of the
SAAO system is slightly different from that of Johnson (Cousins 1984), the difference is very small
(about 0.01 $mag$). The difference therefore between the SAAO system and the Landolt standard is due to the
departure of Landolt's $U-B$ from the Johnson standard system color. Such a large difference may cause problems in
reddening estimates as well as in determining the temperature of hot stars.

As shown in Figures 5 and 6, our $U-B$ scale well matches that of the SAAO scale within 3.6 $mmag$. In addition
there is no systematic difference betweem them. We compare our $U-B$ with Landolt values in the lower two panels
of Figures 8 -- 11. As already confirmed by SB00, we could confirm the systematic differences in our new data
obtained at MAO. The dotted line in $\Delta(U-B)$ versus $B-V$ is from M91, and that in $\Delta(U-B)$ versus
$(U-B)$ is from Bessell (1995). The difference follows the same trends found by M91 and Bessell (1995). But
for $B-V$ = 0.5 -- 1.0 the difference seems to be more negative, while for $B-V > 1.5$ the trend is more
positive. Such a larger difference should be confirmed from more observations.

\begin{figure}[!t]
\epsfxsize=9cm \epsfbox{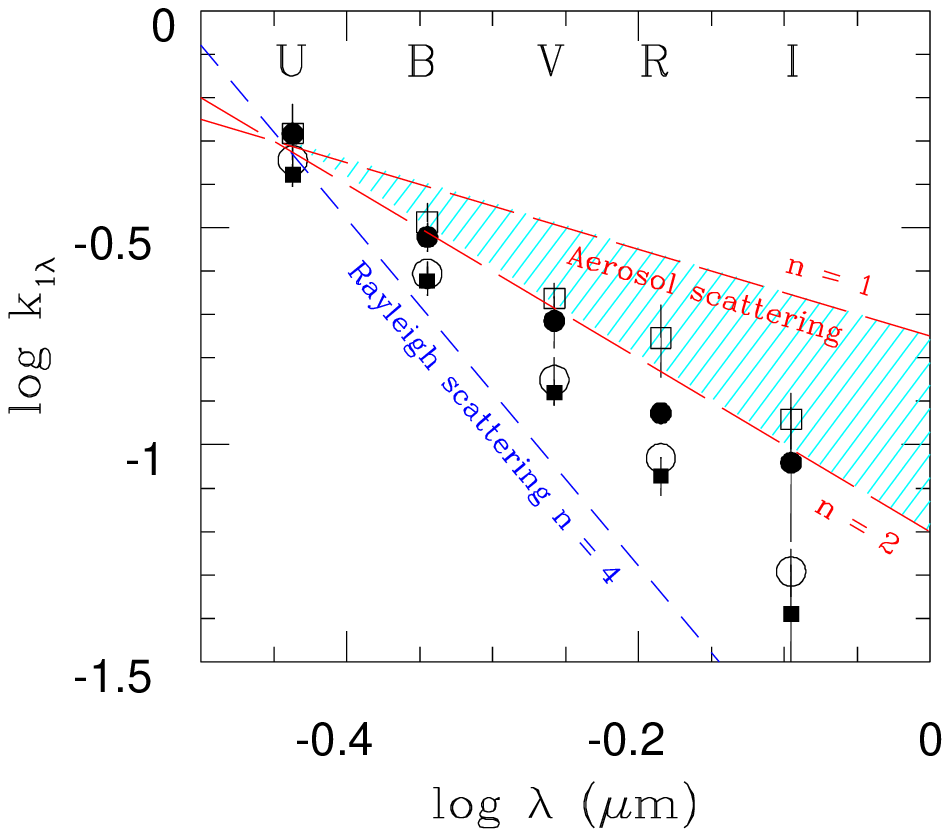} \caption{The relation between wavelength and mean extinction
coefficients for a given season. The shaded area between the two long-dashed lines represents the area affected by
aerosol scattering ($n=1-2$), and the blue short-dashed line denotes the extinction due to pure Rayleigh scattering ($n=$4).
The open square, filled circle, open circle, and filled square represent the mean extinction coefficients
in spring, summer, fall, and winter, respectively.}
\end{figure}

\begin{table}[!t]
\begin{center}
\scriptsize
\bf{\sc  Table 4.}\\
\sc{Seasonal mean value of extinction coefficients} \\
\begin{tabular}{@{}c@{}cccc@{}}
\\ \hline \hline

Season & Spring               & Summer               & Fall                  & Winter \\
\hline
 $U$   & 0.522   $\pm$ 0.015  & 0.520   $\pm$ 0.088  & 0.453   $\pm$ 0.060   & 0.419   $\pm$ 0.012  \\
 $B$   & 0.326   $\pm$ 0.034  & 0.301   $\pm$ 0.022  & 0.248   $\pm$ 0.017   & 0.238   $\pm$ 0.018  \\
 $V$   & 0.217   $\pm$ 0.018  & 0.193   $\pm$ 0.026  & 0.141   $\pm$ 0.018   & 0.132   $\pm$ 0.009  \\
 $R$   & 0.176   $\pm$ 0.033  & 0.118                & 0.093   $\pm$ 0.001   & 0.085   $\pm$ 0.009  \\
 $I$   & 0.115   $\pm$ 0.017  & 0.091   $\pm$ 0.031  & 0.051   $\pm$ 0.006   & 0.041   $\pm$ 0.012  \\
 $n$   & 1.88 $\pm$ 0.09& 2.28 $\pm$ 0.14& 2.75 $\pm$ 0.04& 2.93 $\pm$ 0.10\\
\hline
\end{tabular}
\end{center}
\end{table}

\section {DICUSSION}
\subsection{The Characteristics of Atmospheric Extinction}

We have already shown the yearly variation of the atmospheric extinction coefficients at MAO
in \S $II$. There is an obvious seasonal variation of the primary extinction coefficients.
It is valuable to investigate the main source of extinction.

Atmospheric extinction is mainly caused by scattering of light by air molecules and small
particles swept up in the earth's atmosphere. The scattering efficiency is a function of wavelength.
And therefore the relation between atmospheric extinction coefficients and wavelength of light
can be expressed by

\begin{equation}
k(\lambda) = {\beta \over \lambda^n}
\end{equation}

\noindent
where $k(\lambda)$, $\beta$, and $\lambda$ represent, respectively, the extinction coefficient,
an appropriate constant, and the mean wavelength of filter (Golay 1974). If the extinction
is caused solely by pure air molecules, i.e. Rayleigh scattering, then $n=4$. While if the source
of scattering is due to aerosol pollutants and fine dust, then $n$ is between 1 and 2.

To investigate the seasonal variation of scattering sources we averaged the extinction coefficients
for a given season and present them in Table 4. On 2006, August 22 all extinction coefficients were
abnormally large, and these data were excluded in the average. We also drew the mean
extinction coefficient against wavelength in Figure 12. The mean wavelength for each passband was
obtained from Bessell (1990).
The starting date of each season is March 21, June 22,
September 23, and December $22$, respectively. The calculated power $n$ is also presented in the bottom line of Table 4.

\begin{figure}[!t]
\epsfxsize=9cm \epsfbox{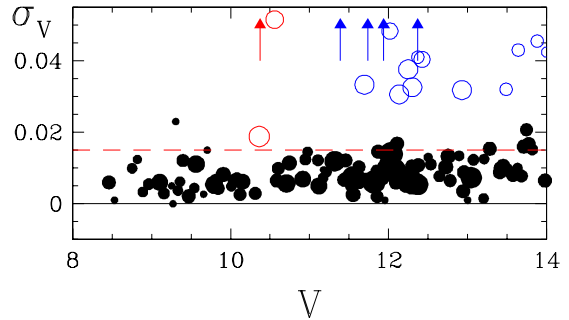} \caption{The $\sigma_{V}$ versus $V$ diagram. The size of circles is proportional
to the number of our observations. The dashed line represents the border between constant stars and variable candidates.
The open circles represent variable star candidates (red circles are stars in M91, while the blue circles and blue arrows
represent stars in L92). }
\end{figure}

Spring is the worst season for observing at MAO and the number of photometric nights was only two.
The resultant power $n \approx 2$ may be biased and therefore less meaningful.

As already shown in Figure 1, the extinction coefficients show a large scatter in summer,
especially in $I$ and $U$. Summer at MAO is famous for its many successively clear days during which
we can easily see dust lanes in the atmosphere.  The summer extinction is consequently strongly affected
by the dust and as a result the power $n$ is about 2, which is smaller than that in fall or winter.

In fall and winter, the mean values of the atmospheric extinction coefficients are smaller and the power $n$ is between
2.7 and 2.9 indicating that Rayleigh scattering is the main source of extinction in these seasons.
This value is very similar to that at SSO (calculated from the mean extinction coefficients in SB00).

\subsection{Variable Star Candidates}

The constancy of standard stars is very important in precision photometry. If variable stars are included in the observations,
larger errors in the extinction coefficients, transformation relations, and photometric zero points are unavoidable. From
repeated observations of standard stars, we found several variable candidates.

\begin{table}[!t]
\begin{center}
\scriptsize
\bf{\sc  Table 5.}\\
\sc{Variable candidates from Menzies et al (1991)} \\
\begin{tabular}{@{}ccccccc@{}}
\\ \hline \hline
 Star & $\Delta V$ & reference &$\sigma_{V}$ & $n$ & CCD & Remark\\
\hline
HD173637   &   0.074 & M91  &   -   &  1  & 2k CCD & V455 Sct   \\
SA109-537  &  -0.019 & M91  & 0.016 &  4  & 2k CCD &    \\
           &  -0.020 & M91  & 0.022 &  8  & 4k CCD &    \\
SA110-353  &   0.039 & M91  &   -   &  1  & 4k CCD &    \\
SA110-441  &   0.022 & M91  & 0.014 &  6  & 2k CCD &    \\
           &   0.003 & M91  &   -   &  1  & 4k CCD &    \\
SA111-1969 &  -0.090 & M91  & 0.023 &  3  & 2k CCD &    \\
           &  -0.046 & M91  & 0.091 & 11  & 4k CCD &    \\
SA111-2009 &  -0.047 & M91  & 0.035 &  3  & 2k CCD &    \\
           &   0.042 & M91  & 0.052 & 10  & 4k CCD &    \\
SA97-345   &  -0.126 & L92  & 0.163 &  3  & 4k CCD &    \\
SA98-733   &  -0.011 & L92  & 0.038 &  7  & 4k CCD &    \\
SA107-456  &  -0.010 & L92  & 0.032 &  7  & 4k CCD &    \\
SA107-458  &  -0.018 & L92  & 0.033 &  7  & 4k CCD &    \\
SA107-459  &  -0.016 & L92  & 0.033 &  7  & 4k CCD &    \\
SA107-602  &  -0.019 & L92  & 0.031 &  7  & 4k CCD &    \\
SA110-266  &   0.000 & L92  & 0.048 &  5  & 4k CCD &    \\
SA110-361  &   0.001 & L92  & 0.040 &  5  & 4k CCD &    \\
SA111-1925 &  -0.067 & L92  & 0.039 &  3  & 2k CCD &    \\
           &   0.020 & L92  & 0.083 &  7  & 4k CCD &    \\
SA111-1965 &  -0.089 & L92  & 0.091 &  3  & 2k CCD &    \\
           &   0.028 & L92  & 0.086 &  6  & 4k CCD &    \\
SA113-260  &   0.036 & L92  & 0.041 &  3  & 4k CCD &    \\
SA113-272  &   0.022 & L92  & 0.045 &  3  & 4k CCD &    \\
SA113-366  &   0.049 & L92  & 0.032 &  3  & 4k CCD &    \\
SA113-372  &   0.039 & L92  & 0.043 &  3  & 4k CCD &    \\
\hline
\end{tabular}
\end{center}
\end{table}

The criteria of identifying variable candidates observed with the 2k CCD are either a large difference in
$V (|\Delta V| \geq 0.05)$
relative to the catalogued value or a larger scatter in $V$ ($\sigma_{V} > 0.015$ if the number of observations were over 3).
We also confirmed the variable candidates observed with the 4k CCD in the same way as for the 2k data for the standard
stars in M91 or L92. These variable candidates are listed in Table 5 and the $\sigma_{V}$ versus $V$ diagram
is shown in Figure 13.

Two known red variables in M91 (BD+1 4774 = BR Psc and GL628 = V2306 oph) were observed only once each, but did not show a
large difference in $V$.
Another known variable HD 173637 (= V455 Sct) did show a large difference in $V$ and is listed in Table 5.
Sixteen variable candidates were found in L92. Two of them (SA111-1925 and SA111-1965) were already detected through their
variability with the 2k CCD. The candidates showed large standard deviations in both data sets. SA97-345 showed very
large standard deviations as well as large differences between our data and the L92 data.
The star was observed three times. In 2006 December 26 the star was fainter by about $0.37 \ mag$ than
on the other days. Other candidates were observed fewer than ten times therefore more observations of these stars are
required to confirm the variability clearly.

\section{SUMMARY}
We observed many standard stars from Menzies et al. (1991) and Kilkenny et al. (1998) at Maidanak Astronomical
Observatory in Uzbekistan to derive the standard transformation relations for the AZT-22 1.5m telescope.
The results from these observations are as follows.

1) We determined the atmospheric extinction coefficients for photometric nights. The primary extinction
coefficients show an evident seasonal variation. We tried to interpret the characteristics of the atmospheric
extinction in the context of scattering sources. In the summer, the extinction seems to be strongly affected by
the dust in the atmosphere, while in fall and winter, Rayleigh scattering due to air molecules seems to be
the dominant source of extinction.

2) We derived the standard transformation relations both for the SITe 2000 $\times$ 800 CCD and the Fairchild 486
4k CCD. Transformation to the standard system is mostly possible with one or two straight lines for all bands except for
$U$ with the Fairchild 486 CCD where we found a non-linear correction term related to the size of the
Balmer discontinuity or the strength of the hydrogen lines.

3) All photometric data were transformed to the SAAO standard system and we found that our data are well
consistent with SAAO zeropoints within a few $mmag$ with a standard deviation of about 0.01 $mag$.
In $U-B$ the standard deviation of the differences was somewhat large (0.02 -- 0.03 $mag$). Such a large scatter
in $U-B$ is very likely due to the SAAO data which was obtained with a small telescope ($\phi=50cm$).

4) We also confirmed the mostly small differences between the SAAO system and the Landolt version. The differences
in $V$, $B-V$, $V-I$, and $R-I$ are systematic, but not very large. But that in $U-B$ is quite large and systematic.
The large difference between standard systems in $U-B$ may cause problems in the determination of stellar parameters
of early type stars.

5) From repeated observations of standard stars we found large differences or a large scatter in $V$ for some
stars. We presented these variable candidates in Table 5.

\acknowledgments

This work is supported by the Astrophysical Research Center for the Structure and Evolution of the Cosmos (ARCSEC \arcsec) at
Sejong University.


\end{document}